\documentclass[a4paper,12pt]{article}

\def\pm   {\partial_{\mu}}

\def\ad   {a^{\dagger}}

\def\HP   {\hat {P}}

\def\al   {\alpha}
\def\be   {\beta}
\def\bet  {\beta}

\def\del  {\delta}

\def\HH {\hat H}

\def\HN {\hat N}

\def\HP {\hat P}
\def\HQ {\hat Q}
\def\HR {\hat R}

\def\HU {\hat U}

\def\HW {\hat W}

\def\HH {\hat H}
\def\heta {\hat\eta}
\def\hetad {\hat\eta^{\dagger}}

\def\HQ {\hat Q}

\def\hw {\hat w}

\def\hrho {\hat{\rho}}

\def\HN {\hat N}

\def\CC {{\cal C}}

{\it}
{\it}
\def\expo {\rm {exp}}{\it}
\def\exp {\rm {e}\it}
\def\tr {\rm {tr}}{\it}
\def\Tr {\rm {Tr}}{\it}
\def\ogar {\overline\gamma_r}

\def\be {\beta}
\def\bet {\beta}
\def\ga {\gamma}
\def\de {\delta}

\def\eps {\epsilon}

\def\sig {\sigma}

\def\la {\lambda}

\def\om {\omega}
\def\Om {\Omega}

\def\hh {{\cal H}}

\usepackage{times}
\usepackage{graphicx}
\begin{document}
\title{ Projected Thermal Hartree-Fock-Bogoliubov approximation in a canonical ensemble.} 
\author{G. Puddu\\
       Dipartimento di Fisica dell'Universita' di Milano,\\
       Via Celoria 16, I-20133 Milano, Italy}
\maketitle
\begin {abstract}
        The Thermal Hartree-Fock-Bogoliubov approximation with reprojection to good quantum numbers
        is analyzed in a canonical ensemble. Simple recipes are given for the evaluation of
        traces, gradients and expectation values in the canonical ensemble.
\par\noindent
{\bf{Pacs numbers}}: 21.60.Jz,$\;\;\;$ 21.10.-k$\;\;\;$ 21.90.+f,$\;\;\;$ 05.30.-d
\vfill
\eject
\end{abstract}
\section{ Introduction.}
         In the past few years huge progresses have been made in the study of
         of nuclei at zero temperature in an ab-initial fashion, using both phenomenological effective
         nucleon-nucleon interactions or effective interaction derived from  more
         fundamental bare nucleon-nucleon interactions (see for example refs. [1],[2] and references in 
         there for recent reviews).
         Unfortunately for nuclei at finite temperature no such advances have been reported so far.
         Most of the studies reported so far in the literature have been made for simplified
         phenomenological interactions. In the Shell Model Monte Carlo approach usually interactions with 
         good sign statistics are used (see for example ref. [3],[4] and references in there).
         Or with  approximate theoretical schemes. 
         For example the static path approximation, although formulated for general interactions
         has been applied only for quadrupole-quadrupole or pairing+quadrupole phenomenological
         interactions (ref. [5] but not to the more complicated realistic effective 
         interactions such as
         FPD6 or GXPF1A or similar, or to effective interaction derived from the more
         fundamental bare nucleon-nucleon interactions. This is not very surprising since
         even at the level of the static path approximation the inclusion of all multipole
         parts of the interaction leads to statistical density operators that are not hermitian
         and therefore the evaluation of the partition function can be rather difficult, if possible.    
         Surprisingly, even the Hartree-Fock-Bogoliubov with the restoration of all quantum
         numbers (cf. ref. [6] for a recent reformulation) has never been applied at finite
         temperature with effective interactions (whether realistic, that is fitted to the experimental
         data, or derived from nucleon-nucleon interactions with renormalization methods).
         The purpose of this paper is to analyze some of the  difficulties
         associated with the thermal HFB approximation (THFB) and to propose a simple method to work
         directly in a canonical ensemble (that is all traces are taken in the subspace
         of the full Hilbert space that has a specified total number of particles).
         The use of quantum-number projected statistical density operators is very relevant since they
         smooth out sharp phase transitions (ref.[7]).
         This paper is organized as follows. In section 2, we briefly review the projected
         HFB approximation at finite temperature. In section 3 we summarize the main
         properties of the statistical density operators,  discuss
         the methods needed to evaluate traces in a canonical ensemble as well as
         gradients (with respect to the variational parameters) and expectation values.
         In section 4 we present some concluding remarks.
\bigskip
\section {Recap of the projected THFB approximation.}
         There are two methods to obtain the THFB with restoration of exact quantum numbers.
         One method consists in the  minimization of the free-energy defined as
$$
F=<\HH>-T S
\eqno(1)
$$
 where $\HH$ is the many-body Hamiltonian, and 
$$
<\HH>=Tr(\hw \HH)
\eqno(2)
$$
          with
         $\hw$ being the statistical operator
$$
\hw= \HP \exp^{-\beta \HH_0}\HP/Z
\eqno(3)
$$
         T is the temperature
         $T=1/\beta$, $\HH_0$ is a trial quadratic Hamiltonian in the creation and annihilation 
         operators that breaks all symmetries, including
         conservation of particle number, $\HP$ is the projector that restores good quantum numbers, and 
$$
Z=Tr(\HP \exp^{-\beta \HH_0})
\eqno(4)
$$
         is the projected partition function for this trial Hamiltonian. $S$ is the entropy defined as
$$
S= -Tr(\hw \ln(\hw))
\eqno(5)
$$
        The trace operation is taken over the full Hilbert space. The minimization is carried out
        on the trial Hamiltonian $\HH_0$.
        The difficulty with this approach is the presence of the projector $\HP$ in the logarithm.
        It can be remedied by replacing, as  an approximation (cf. ref.[6]),  
        the entropy with $\beta Tr(\hw \HH_0)+\ln Z$.  
        Despite that, the task is not simple, since
        we need to minimize a functional we also need the gradients of this functional
        with respect to the variational parameters, i.e. the matrix elements appearing in the
        trial Hamiltonian $\HH_0$.
        The true minimum for $F$ is obtained when $\HH_0$ is the exact many-body Hamiltonian and
        in this case 
        $F[\HH]=-{1\over \beta} Tr(\HP \exp^{-\beta \HH})$.
\par
       As an alternative approach we can approximate the propagator $\exp^{-\beta \HH}$ appearing in 
       the free-energy minimum, with
       some propagator that at the mean-field level reproduces the HFB approximation.
       We can then make use of the approach proposed almost 30 years ago in ref. [8], based
       on a formulation of the functional integral expression of the partition function
       different from the one obtained using the Hubbard-Stratonovich transformation.
       As well known, functional integrals for the partition function obtained
       by approximating infinitesimal evolution operators $\exp^{-\eps \HH}$ using  the
       Hubbard-Stratonovich transformation do not lead, in the mean-field approximation,
       to the Hartree-Fock or to the Hartree-Fock-Bogoliubov approximation. In ref.[8]
       it has been suggested that adding appropriate terms in the elementary propagators appearing
       inside the functional integral, that do not change the functional integral, but
       that crucially change the mean-field, one can obtain either the
       Hartree-Fock or the Hartree-Fock-Bogoliubov approximation (cf. refs. [8],[9] for a detailed
       discussion). In ref.[8] only the unprojected partition function has been discussed.
       However it enables to define which terms must be added inside the functional
       integral in order to obtain the HFB approximation  by maximizing the integrand which 
       contains
       only one trace (in contrast with several traces of the previous method). 
       We shall rewrite here the functional that needs to be maximized. We use notations
       close to the ones in ref. [8]. $i,j,k,l,..$ denote single-particle states.
       The effective Hamiltonian is written as 
$$
\HH=\sum_{ij}<i|K|j>\ad_i a_j+{1\over 2}\sum_{ijkl}<ij|V|kl>\ad_i\ad_j a_l a_k
\eqno(6)
$$
       $K$ is the one-body term that also includes the chemical potential term $-\mu_F$ (for
       simplicity we do not distinguish neutrons and protons) and $V$ is the two-body interaction.
       Let us define the operators
$$
\hrho_{ij}=\ad_i a_j,\;\;\;\;\;\;\hetad_{ij}=\ad_i \ad_j,\;\;\;\heta_{ij}=a_j a_i
\eqno(7)
$$
       and the following matrix multiplication rules
$$
\sig v \rho=\sum_{ijkl} \sig_{ij}v_{ijkl}\hrho_{kl}
\eqno(8)
$$
$$
\mu u \hetad=\sum_{ijkl} \mu_{ij}u_{ijkl}\hetad_{kl}
\eqno(9)
$$
$$
\nu w \heta=\sum_{ijkl} \nu_{ij}w_{ijkl}\heta_{kl}
\eqno(10)
$$
       then the THFB approximation is obtained by maximizing $\exp^{-\beta \Om}$ where
$$
\Om = {1\over 2}\sum_{ij}(\sig_{ij}^2+\mu_{ij}^2+\nu_{ij}^2)-{1\over \bet} \ln Tr(\HU)+\mu_F A
\eqno(11)
$$
      where
$$
\HU=\exp^{-\bet \hh}
\eqno(12)
$$
     and
$$
\hh= \sum_{ij}<i|K|j>\ad_i a_j-\sig v \hrho-(\mu+i \nu)w\hetad-(\mu-i \nu)u\heta
\eqno(13)
$$
     Note here that the trace operation does not include the projectors. Also we have replaced $u^{\star}$
     in ref. [8] with $w$ for more generality.
     In order to obtain the HFB approximation (without the projectors) one has to
     fix $v,u,w$ such that
$$
(\tilde v v)_{ijkl}= -V^A_{ik,j,l},\;\;\;\;(\tilde w u)_{ijkl}=-{1\over 8} V^A_{ijkl}
\eqno(14)
$$
    (note the peculiar order of the indices in the first equation) where $V^A$ is the antisimmetrized
    interaction.
    This is the only result we need from ref.[8].
    A simple way to determine the matrix $v$ is the following. Define first the indices 
    $\al=(i,j) $ and $\beta=(k,l)$ and the real hermitian matrix $W_{\al,\bet}=V^A_{ikjl}$. 
    Then
    decomposing $W$ into its eigenmodes we have $W= X \om \tilde X$ where $\om$ is diagonal.
    Therefore one can make the choice $ v= \sqrt{-\om} \tilde X$. Similarly we can decompose
    $V^A_{(ij)(kl)}$ (with this grouping of the indices) in its eigenmodes $V^A= Y \sig \tilde Y$ with 
    $\sig$ being diagonal, and 
    take
    $ u=w=\sqrt{-{1\over 8}\sig}\tilde Y$. Since the eigenvalues in both cases can have any sign
    $u,v,w$ are generally complex and therefore  the Hamiltonian $\hh$ of eq.(13) is not
    necessarily hermitian, except at the stationary point.
\par
    It is sensible to consider the following projected functional
$$
\Om_P = {1\over 2}\sum_{ij}(\sig_{ij}^2+\mu_{ij}^2+\nu_{ij}^2)-{1\over \bet} \ln Tr(\HP\HU)+\mu_F A
\eqno(15)
$$
    where $\HU$ is given by eqs.(12)-(13). Once this functional has been minimized with
    respect to $\sig_{ij},\mu_{ij},\nu_{ij}$ one can evaluate whatever thermal averages
    one wishes to compute. Hence the central task is to evaluate as efficiently as possible
    $\ln Tr(\HP\HU)$ and its gradients with respect to $\sig_{ij},\mu_{ij},\nu_{ij}$. Once this is done
    we can minimize $\Om_P$ using for instance quasi-newtonian minimization techniques (ref.[10]).
    In the following section we shall discuss precisely how can one evaluate efficiently
    $\ln Tr(\HP\HU)$ and its gradients. It will be shown that actually one can work directly
    in a canonical ensemble rather than in the grand-canonical one.
    As a final comment, note that the free energy obtained in this way may not
    be optimal in the sense of the first method (this is the reason why we have used a different symbol), 
    but at least we avoid entirely the problem of evaluating the logarithm of a projected operator.
    Also, since $ \Om_P$ is complex, it is best to minimize its real part and
    imposing a penalty function to suppress its imaginary part. That is 
    it is better to minimize the functional $ G=Re(\Om_P)+ c (Im(\Om_P))^2$, $c$ being
    a positive number that controls the strength of the penalty function. 
\par 
    The strategy of approximating the exact propagator with an approximate one, and then
    evaluating its associated free energy, is not a new one. Also ref. [5]
    where the propagator was replaced by the projected SPA propagator, follows the
    same line, instead of minimizing the functional of eq. (1). 
\bigskip
\section {Technique for the evaluation of $\ln Tr(\HP\HU)$, its gradients and expectation values 
         in a canonical ensemble.}
\bigskip
       Let us first rewrite the propagator $\HU=\expo(-\bet \hh)$ in a more convenient form as
$$
\HU=\expo(\CC)\;\; \expo( {1\over 2} \ogar S \ga_c)\equiv \expo(\CC) \HW
\eqno(16)
$$
      where $\ga_c= col(a,\ad)$ is the column vector of all creation and all annihilation operators
      for all $N_s$ single-particle states (including the isospin label), and $\ogar=row(\ad,a)$.
      The matrix $S$ is a $2N_s\times 2N_s$ matrix of the type
$$
S= \left(\begin{array}{cc}S_{11} & S_{12}\\ S_{21} & S_{22} \end{array}\right )
\eqno(17)
$$
      with
$$
S_{11}=-\bet(K-\sig v),\;\;\;S_{12}= 2\bet (\mu+i\nu)w,\;\;\;S_{21}= -2\bet(\mu-i\nu)u
\eqno(18)
$$
      and $S_{22}=-\tilde S_{11}$. The matrices $S_{21}$ and $S_{12}$ are skew-symmetric.
      Moreover $\CC=\sum_i (S_{11})_{ii}$. 
      The most generic projector has the following structure
$$
\HP=\sum_\om D^{\star}(\om) \exp^{\ad r a}
\eqno(19)
$$
      where $\om$ is a set of rotation angles (for example the Euler angles and/or the fugacities)
      and $D^{\star}$ ia function of the rotation angles (for example the Wigner functions
      in the case of angular momentum projection) and $r$ is a matrix in the single-particle
      indices. Therefore we will be concerned to the evaluation of quantities of the type 
      $Tr(\exp^{\ad r a}\HW )$. In the following subsections we first recall briefly some
      of the properties of products of operators like $\HW$, called statistical density operator (SDO)
      which are not necessarily hermitian, and then we will apply them
      to the evaluation of the traces and their gradient with respect to the matrix elements
      of $S$
\bigskip
\par\noindent{\it{$3.1 $ Basic properties of Statistical Density Operators.}}
\bigskip
\par
      For a thorough discussion we refer the reader to ref.[11]. The most useful properties
      of the SDO's are the following.
\par\noindent
      1. Group property. The product of two SDO's is a SDO with some matrix $S$ associated
      to it.
\par\noindent
      2. To any SDO there is a matrix associated to it. More precisely if
$$
\HW = \expo( {1\over 2} \ogar S \ga_c)
\eqno(20)
$$
      then the associated matrix is
$$
W  =  \exp^S
\eqno(21)
$$
      (without the caret), which preserves the group structure, that is, if $\HW=\HW_1\HW_2$ then
      $W=W_1 W_2$. 
\par\noindent
      3. The matrix $W$ satisfies the identity $W^{-1}=\eta \tilde W \eta$ where 
      $\eta = \left(\begin{array}{cc}0 & 1\\ 1 & 0 \end{array}\right ) $.
\par\noindent
      4. The matrix $W$ can be diagonalized by a similarity transformation of the same type $T$, that 
     satisfies $T^{-1}=\eta \tilde T\eta$ (cf. ref. [12] for a proof).
\par\noindent
      5. The eigenvalues of the matrix $S$ come in opposite pairs 
       $(\la_1,\la_2,..\la_{N_s},-\la_1,-\la_2,..-\la_{N_s})$.
        Moreover the Grand-Canonical trace of $\HW$ is given by (cf.ref.[12])
$$
Tr_{gc} \HW = \prod_{k=1}^{N_s} ( \exp^{\la_i/2} + \exp^{-\la_i/2} )
\eqno(22)
$$
\par\noindent
      6. Taking the square of eq.(22) we have
$$
( Tr_{gc} \HW )^2=\det(1+W)
\eqno(23)
$$
\par\noindent
      7. Any SDO can be written as a product of three special SDO's (ref.[11]), i.e.
$$
\HW = \expo({1\over 2} \ad C \ad)  \expo( {1\over 2}\ogar \left(\begin{array}{cc}Y & 0\\ 0 & -\tilde Y 
\end{array}\right )\ga_c)  \expo({1\over 2} a D a)     
\eqno(24)
$$
      with
$$
 C = W_{12} W_{22}^{-1},\;\;\; D= W_{22}^{-1} W_{21},\;\;\; \exp^{-Y}=\tilde W_{22}
\eqno(25)
$$
      $W_{12},W_{21},W_{22}$ are the $N_s\times N_s$ blocks of $W$. Because of the
      relation $\tilde W \eta W\eta =1$, the matrices $C$ and $D$ are skew-symmetric.
\par
      The following properties are more relevant to the topic under discussion and can be proved using the 
      above. 
\par\noindent
      8. The vacuum expectation values of $\HW$ is given by $\det(W_{22})^{1/2}$ and the vacuum
      expectation value of the product $\HW(1) \HW(2)$ is given by
$$
<0|\HW(1) \HW(2)|0>=\det(W_{22}(1))^{1/2} \det(W_{22}(2))^{1/2}\prod_k {'}(1+\nu_k)
\eqno(26)
$$
      where the $\nu_k$ are the eigenvalues of $D(1) C(2)$ which come in degenerate pairs
      and the ' denotes one eigenvalue per degenerate pair.
\par\noindent
      9. The grand canonical trace of $\exp^{\al\HN}\HW$  can be rewritten using the eigenvalues $\mu_k$ of
      the following matrix
$$
 M  = \left(\begin{array}{cc}\tilde W_{22}^{-1} & C\\ -D & W_{22}^{-1} \end{array}\right ) 
\eqno(27)
$$
      which come in degenerate pairs $(\mu_1,\mu_2,..,\mu_{N_s},\mu_1,\mu_2,..,\mu_{N_s})$,
      as
$$
\Tr_{gc}(\exp^{\al\HN}\HW)= \det(W_{22})^{1/2} \prod_{k=1}^{N_s}(1+ z \mu_k)
\eqno(28)
$$
      where $z=\exp^{\al}$, and the product includes  one eigenvalue per degenerate pair.
      This very important property has been proved in a slightly different form in ref. [12].
      The only difference here is the use of $M=S_p S_v^{-1}$, instead of $M= S_v^{-1} S_p$ 
      (cf. ref.[12] for the details and the definitions of these matrices).
      In the next subsection we will discuss
      how the above equation allows us to work directly in a canonical ensemble rather than
      in the grand-canonical one.
\bigskip
\par\noindent{\it{$3.2 $ Evaluation of traces.}}
\bigskip
\par
      Let us discuss first some of the implications of eq. [28].
      Some of these have been discussed in ref. [12]. First if we know the matrix $S$ that defines
      the SDO we can unambiguously determine the phase of $\det(W_{22})^{1/2}$. In fact 
      we can set $z=1$ in eq. (28) (that is $\al=0$) and obtain unambiguously $\det(W_{22})^{1/2}$
      using eq. (22) and eq. (28).
\par\noindent
      Second, once we have done this, we can isolate the coefficient of $z^N$ in the product appearing in 
      eq.(28), $\xi_N$, and therefore
      the canonical ensemble trace for $N$ particles of $\HW$ is given by
$$
\Tr_c (\HW)=\xi_N {\prod_{k=1}^{N_s} ( \exp^{\la_i/2} + \exp^{-\la_i/2} )
\over \prod_{k=1}^{N_s}(1+\mu_k) }
\eqno(29)
$$
      The expression for the coefficient $ \xi_N$ is the following
$$
\xi_N=\sum_{i_1<i_2<...<i_N} \mu_{i_1}\mu_{i_2}...\mu_{i_N}
\eqno(30)
$$
      These coefficients can be constructed iteratively by defining the 
      coefficients $\xi(n,s)$ for $n$ particles using the first $s$ distinct eigenvalues
$$
\xi(n,s)= \xi(n,s-1)+\mu_s \xi(n-1,s-1)
\eqno(31)
$$
      There is a nontrivial consequence of eq. (29). Qualitatively, the $M$ matrix is scale independent.
      In order to see this,
      let us recall that the eigenvalues $\la$ are proportional to the temperature. Let $T$ be the
      matrix of the eigenvectors of $S$ that is
$$
 S  = T  \left(\begin{array}{cc}
 \la & 0\\ 0 & -\la\end{array}\right) T^{-1}
\eqno(32)
$$
      (cf. property 5.). We can order the eigenvalues so that the first $N_s$ have positive
      real part.  
      Let us call $T_{\al,\be}$ and $I_{\al,\be}$ the partitions of $T$ and $T^{-1}$ ($\al,\be=1,2$).
      Also set $D= \exp^{\la}$ and $ d=\exp^{-\la}$ which represent the large and the small scale. 
      Then we have
$$
W_{22}= T_{21} D I_{12} + T_{22} d I_{22} = L_{22} + S_{22}
\eqno(33)
$$
$$
W_{12}= T_{11} D I_{12} + T_{12} d I_{22} = L_{12} + S_{12}
\eqno(34)
$$
$$
W_{21}= T_{11} D I_{11} + T_{22} d I_{12} = L_{21} + S_{21}
\eqno(35)
$$
      which shows the large and small scale ( arrays $L$ and $S$) of the various blocks of $W$.
      One can write
$$
 D = (1+ L_{22}^{-1} S_{22})^{-1} L_{22}^{-1} L_{21}(1+L_{21}^{-1} S_{21})
\eqno(36)
$$
      Using the definitions in eqs.(33)-(35) we have $L_{22}^{-1} L_{21}= I_{12}^{-1} I_{11}$.
      therefore, especially at low temperature, $D$ is scale (i.e. temperature) independent.
      Similarly for $C$ we have
$$
C = (1+S_{12} L_{12}^{-1}) L_{12} L_{22}^{-1} (1+ S_{22}L_{22}^{-1})^{-1}
\eqno(37)
$$
      Again we have $L_{12} L_{22}^{-1}= T_{11} T_{21}^{-1}$ which is temperature independent.
      The remaining factors in $C$ and $D$ have a mild temperature dependence.
      Therefore the matrix $M$ is roughly scale independent.
      Hence we arrive at the conclusion that the temperature dependence of the canonical ensemble
      trace of $\HW$ is mostly in the vacuum expectation values ( the ratio in eq. (29) ).
      We obtained this conclusion under the assumption that the inverses of $T_{\al,\be}$ and $I_{\al,\be}$
      exist. This is not true in general. That is, the various block can be singular even though
      the $T$ and $I$ matrices are not. Despite this limitation in the proof, it is a surprising result.
      This scale independence of the the eigenvalues of $M$ has been studied in detail in ref. [13]
      although only for the pure pairing model.
\par\noindent
      This nearly temperature independence  implies that at low temperature $\xi_N$ does not contribute 
      to the energy at low temperature.
      This in turn implies that any dependence of the energy from the number of 
      particles must come from the vacuum
      contribution. Differently stated, the chemical potential is the only parameter that determines
      the dependence of the energy on the number of particles. The chemical potential, which plays an 
      essential role in the grand-canonical ensemble, retains its importance also in the canonical 
      ensemble.
      Being a free parameter in the canonical ensemble, and since we seek to minimize the free energy 
      functional, it must be fixed in order to minimize the free 
      energy, hence $\pm \Om= 0$ with $\Om$ given by eq. (15).
      We have assumed that the eigenvalues of $M$ are dominated by the $C$
      and $D$ blocks. If, as in the case of lack of pairing, these matrices are zero,  the
      matrix $W_{22}^{-1}$ which appears in $M$ will be relevant in the determination of
      the energy and this matrix has an exponential dependence on the temperature. 
\par
      So far we have discussed the evaluation of the trace in the canonical ensemble, without
      any other projector. Let us turn now to the evaluation of
      $\Tr_c (\expo(\ad r a)\HW)$. Note that $<0| \expo(\ad r a)|0>=1$.
      Let us first rewrite 
$$
 \expo(\ad r a)= \expo({1\over 2} \sum_i r_{ii})
\expo( {1\over 2}\ogar \left(\begin{array}{cc}r & 0\\ 0 & -\tilde r
\end{array}\right )\ga_c) \equiv \expo({1\over 2} \sum_i r_{ii}) \HR
\eqno(38)
$$
     and make use of property 8. The result for the vacuum contribution is $\det(W_{22})^{1/2}$
     as in the case $r=0$. The associated matrix to 
$$
\HW' = \HR \HW 
\eqno(39)
$$
     is
$$
W' = \left(\begin{array}{cc}\exp^r & 0\\ 0 & \exp^{-\tilde r}
\end{array}\right ) W
\eqno(40)
$$
     evaluating the matrix $M'$ of $W'$ (cf. eq.[27]) we obtain that the eigenvalues 
     of $M'$ are the same of the matrix 
$$
M^{(r)}= \left(\begin{array}{cc}\exp^r & 0\\ 0 & \exp^{\tilde r}
\end{array}\right ) M
\eqno(41)
$$
     where $M$ is relative to $W$ only. The final result for the canonical trace in this case is
$$
\Tr_c (\expo(\ad r a)\HW)= \det(W_{22})^{1/2} \xi_N(\mu^{(r)})
\eqno(42)
$$
     where the $\mu^{(r)}$ are the eigenvalues of $M^{(r)}$, and $\xi_N(\mu^{(r)})$ is given by the 
     recursion relation of eq.(31) with the "rotated" eigenvalues $\mu^{(r)}$. 
     The vacuum contribution $\det(W_{22})^{1/2} $ is again given by
$$
\det(W_{22})^{1/2}= {\prod_{k=1}^{N_s} ( \exp^{\la_i/2} + \exp^{-\la_i/2} )
\over \prod_k^{N_s}(1+\mu_k) }
\eqno(43)
$$
     Where the $\mu_k$'s are the unrotated eigenvalues of $M$.
     We stress again that the eigenvalues of $M^{(r)}$ are two-fold degenerate and that
     only one eigenvalue per pair must be taken in the evaluation of $\xi_N$.
\par
     From a numerical point of view, the determination of the eigenvalues $(\la_k,-\la_k)$
     of $S$ (cf. eq.(32)) does not pose a problem. The determination of the eigenvalues
     $(\mu_k,\mu_k)$ and $(\mu^{(r)},\mu^{(r)})$ does. As discussed previously, the partitions
     of $W$ carry both a large and a small scale that cancel out in the determination
     of the arrays $C$ and $D$. This analysis however is based on the existence of the inverses
     of the partitions of $T$ and of its inverse, which cannot be guaranteed. The most reliable
     way to prevent loss of accuracy in the determination of $M$  for arbitrarily
     large values of $\bet$ is the following. Let us divide the interval $[0,\bet]$ in $N_{\bet}$ equal
     intervals and  let us define the matrix $u$ associated to the propagator
     in each interval. The $c$ and $d$ (we use small letters for each interval) matrices can be
     determined without loss of accuracy. In force of the group property of the propagators we
     have (right to left propagation)
$$
W(n)= u W(n-1)
\eqno(44)
$$
     where $n=2,..N_{\bet}$ and $W(1)=u$. We seek a "propagation" law for $C$ and $D$ as the interval
     is enlarged. Define the following auxiliary matrices for each elementary interval
$$
\gamma= u_{11}^{-1}u_{12},\;\;\;\;\delta = u_{21}u_{11}^{-1}
\eqno(45)
$$
     Then we have
$$
C(n) = u_{11} [ C(n-1)+\gamma ] [ 1+ d C(n-1) ]^{-1} u_{22}^{-1}
\eqno(46)
$$
     At each step, $C(n)$ is well determined since the matrix $u$ is close to $1$.
     For the determination of $D$ we found more convenient to propagate $W$ from the left to
     right, i.e.
$$
W'(n)= W'(n+1) u,\;\;\;\;(n=N-1,N-2,..1)
\eqno(47)
$$
     with $W'(N_{\bet})=u$, using the definition of $D$ we find
$$
D(n)= u_{22}^{-1} [ 1+D(n+1) c ]^{-1} [D(n+1) +\delta ] u_{11}
\eqno(48)
$$
     with $D(N_{\bet})=d$. Again in the backward propagation, $D$ is numerically stable. 
     There is a final point which must be discussed, that is whether $u_{22}$ has an inverse.
     If we were working with hermitian $S$ the Bloch-Messiah theorem states that indeed, $W_{22}$
     can have zero eigenvalues. These however would be associated to empty single-particle states,
     which are unlikely to happen, since the preferred basis is the harmonic oscillator basis.
\par
     The remaining task is to give some prescriptions for the determination of the gradients
     of $\Tr_c (\HR\HW)$ with respect to all  matrix elements 
     $S_{ij}$. It is simpler to evaluate the variations of $\ln(\Tr_c (\HR\HW))$, and the result is
$$
\delta(\ln(\Tr_c (\HR\HW)))=\sum_{k=1}^{N_s} {1-\exp^{-\la_k} \over 1+\exp^{-\la_k}}{1\over 2}
\delta(\la_k)-\sum_{k=1}^{N_s} {1\over 1+\mu_k}\delta(\mu_k)+\sum_{k=1}^{N_s} f_k(\mu^{(r)})
\delta(\mu^{(r)}_k)
\eqno(49)
$$
     where the fractional partition function (fpf) $f_k$
$$
f_k(\mu^{(r)})= {{\partial (\ln(\xi(\mu^{(r)} ))} \over{ \partial \mu_k^{(r)}} }
\eqno(50)
$$
     satisfy the sum rule $\sum_{k=1}^{N_s}\mu_k f_k = N$ (the number of particles).
     These fpf $f_k$ can be obtained from the $\mu'$ by excluding the eigenvalue $\mu_k$
     and decreasing the number of particles by one.
     The various $\delta \la_k$, $\delta \mu_k$ etc. can be evaluated using first order perturbation 
     theory.
     Perturbation theory on $M$ or $M^{(r)}$ is a bit involved. The simplest way is first to determine
     $\de W_{ij}$ at each interval . To simplify the notations let us call with greek letters the 
     eigenvalue indices of
     $S$ and let $T$ be the matrix that diagonalizes $S$. Using the sum convention over
     repeated indices, we have
$$
\de W_{ij}= T_{i\al} \de W_{\al\bet} T_{\bet j}^{-1}
\eqno(51)
$$
$$
\del W_{\al\bet}= F_{\al\bet} \de S_{\al\bet}\;\;\;\;{\rm {no \; sum}}
\eqno(52)
$$
$$
\de S_{\al\bet}= T_{\al r}^{-1} \de S_{rs} T_{s \bet}
\eqno(53)
$$
     where
$$
F_{\al\bet}= {\exp^{\la_{\al}}-\exp^{\la_{\bet}} \over \la_{\al} -\la_{\bet}},\;\;\;\;\;(\la_{\al}\not= 
\la_{\bet})
\eqno(54)
$$
     and  $F_{\al\al}=\exp^{\la_{\al}}$. The matrix $\de S_{rs} $ has only two non-zero elements. 
     We have performed a numerical comparison of this method to evaluate the gradients with a simple-minded 
     approach that uses the numerical evaluation of the derivatives and found that it is numerically 
     stable even for large values of $\beta$ for not too large $N_{\bet}$.
\bigskip
\par\noindent{\it{$3.3$  Evaluation of one-body and two-body expectation values.}}
\bigskip
\par

     Let us now turn to the problem of evaluating canonical ensemble traces involving one-body and two-body
     observables. Consider the operator 
$$
\HQ(\eps)=\expo( \eps \ad q a)
\eqno(55)
$$
     where $ \ad q a= \sum_i \ad_i q_{ij} a_j$, and let us evaluate
$$
Y_c(\eps) = \Tr_c (\HQ \HR \HW)\equiv Y_0 + \eps Y_1 + {1\over 2} \eps^2 Y_2+...
\eqno(56)
$$
    where the subscript $c$ stands for canonical, up to second order in $\eps$, The second order term gives 
    the matrix expectation values of
    $(\ad q a)^2$. $\ad q a$ could be for instance one of the eigenmodes of $\ad_i a_l v(ij,kl) \ad_j a_k$
    in terms of the pairs $(i,l)$ and $(j,k)$.
    Let $\HR$ be the "rotation operator" that appears in the remaining projectors.
    Consider 
$$
Y_{gc}(z,\eps) = \Tr_{gc} (\exp^{\alpha \HN} \HQ \HR \HW)\equiv Y_{gc 0}+\eps Y_{gc 1}+{1\over 2}\eps^2 
Y_{gc 2}
\eqno(57)
$$
    where $z=\exp^{\al}$. We simply have to determine the coefficient of $z^N$ in $Y_{gc}(z,\eps)$
    up to second order in $\eps$ for $N$ particles.
    From eq.(42), we have
$$
Y_{gc}(z,\eps) = |W_{22}|^{1/2} \prod_{i=1}^{N_s} (1 +z\mu_i^{''})
\eqno(58)
$$ 
     where the product contains the eigenvalues (one per degenerate pair) of
$$
M^{''} = \left(\begin{array}{cc} QR & 0\\ 0 & \tilde R\tilde Q 
\end{array}\right ) M
\eqno(59)
$$
   where $R= \exp^r$. The matrix $M^{''}$ can be rewritten as 
$$
M^{''} = \left(\begin{array}{cc} 1 & 0\\ 0 & \tilde R \end{array}\right )
      \left(\begin{array}{cc} Q & 0\\ 0 & \tilde Q \end{array}\right )
      \left(\begin{array}{cc} 1 & 0\\ 0 & R^*\end{array}\right )M^{(r)}
\eqno(60)
$$
     using the fact that $R$ is unitary in the cases of physical interest.
     In order to simplify the notations let us set
$$
M^{''} = M^{(r)}+ \delta M^{(r)},\;\;\;\;\;\; \delta=\eps\delta_1+{1\over 2}\eps^2\delta_2
\eqno(61)
$$
     and 
$$
\eta = (1+zM^{(r)})^{-1} \delta M^{(r)}
\eqno(61)
$$
    Then, expanding the determinant $|1+z M^{''}|^{1/2}$ up to second order in $\eta$ we obtain
$$
|1+z M^{''}|^{1/2} = |1+z M^{(r)}|^{1/2} ( 1+ {1\over 2} \tr (\eta) +{1\over 8} \tr^2(\eta)-{1\over 4} 
\tr(\eta^2)
\eqno(62)
$$
    Using the diagonal representation of $M^{(r)}$ and the fact that $ Y_{gc}$ must be
    a polynomial in $z$ we obtain, after some algebra
$$
Y_{gc 1} =|W_{22}|^{1/2} \sum_j \left (\prod_{i\not= j} (1+z \mu_i^{(r)}) \right ) z \mu_i^{(r)} [ 
(\de_1)_{jj}+(\de_1)_{\overline j\overline j}  ]
\eqno(63)
$$
    and
$$
Y_{gc 2}= |W_{22}|^{1/2}\big [ \sum_{i<j} \left (\prod_{k\not= i,k\not= j} (1+z \mu_k^{(r)}) \right ) 
   z^2 \mu_i^{(r)}\mu_j^{(r)} G_{ij} + 
$$
$$
\sum_j \left (\prod_{i\not= j} (1+z \mu_i^{(r)}) \right ) z \mu_i^{(r)} [
(\de_2)_{jj}+(\de_2)_{\overline j\overline j} \big ]
\eqno(64)
$$
    where 
$$
G_{ij}= {1\over 2} [(\de_1)_{ii}+(\de_1)_{\overline i\overline i}][(\de_1)_{jj}+(\de_1)_{\overline 
j\overline j} ]-
[(\de_1)_{ij}(\de_1)_{ji}+(\de_1)_{i \overline j}(\de_1)_{\overline j i}+
(\de_1)_{\overline i j}(\de_1)_{j \overline i} + (\de_1)_{\overline i\overline j}
 (\de_1)_{\overline j\overline i} ]
\eqno(65)
$$
     In eqs. (63),(64), sum and product are performed considering one member per degenerate pair and
     $\overline i$ represents the state which has the same eigenvalue of the state labeled by $i$.
     Since terms containing $1/(1+z \mu_i^{(r)})^2$ must cancel out, we find 
     $(\de_1)_{ii}=(\de_1)_{\overline i\overline i}$ and $(\de_1)_{i \overline i}=0$.
     We can now proceed directly to the canonical ensemble and define in addition to the fractional
     partition function $f_i$ of the previous subsection, the fractional partition function for a pair
     $(ij)$
$$
f^{(2)}_{ij}= {1\over \xi_N(\mu^{(r)})} \xi_{N-2}\big (\mu^{(r)}\not=(\mu_i^{(r)},\mu_j^{(r)}) \big )
\eqno(66)
$$
     Appropriately $ \mu_i^{(r)} f_i$ and $\mu_i^{(r)}\mu_j^{(r)} f^{(2)}_{ij}$ can be interpreted as 
     the occupation numbers and pair occupation numbers of nucleons.The quantities
     $\mu_i^{(r)}\mu_j^{(r)} f^{(2)}_{ij}$ satisfy the sum rule $\sum_{i<j} \mu_i^{(r)}\mu_j^{(r)} f^{(2)}_{ij}=N(N-1)/2$.
     The final results for one-body and two-body canonical ensemble traces are as follows
$$
\Tr_c(\ad q a \HR\HW)= \tr_c(\HR\HW)\sum_k \mu_k^{(r)}f_k(\mu^{(r)}) ((\de_2)_{jj}+(\de_2)_{\overline 
j\overline 
j} )
\eqno(67)
$$
     and
$$
\Tr_c((\ad q a)^2 \HR\HW)=\tr_c(\HR\HW)[\sum_i \mu_i^{(r)}f_i  [
(\de_2)_{jj}+(\de_2)_{\overline j\overline j} \big ]+
\sum_{i<j}\mu_i\mu_j f^{(2)}_{ij} G_{ij} ]
\eqno(68)
$$
     The first term can be eliminated by rewriting $(\ad q a)^2 =\ad q^2 a +\ad_i\ad_j q_{il}q_{j,k}a_ka_l$,
     hence     
$$
\Tr_c( \ad_i\ad_j q_{il}q_{jk}a_ka_l \HR\HW)=\tr_c(\HR\HW)\sum_{i<j}\mu_i^{(r)}\mu_j^{(r)} f^{(2)}_{ij} 
G_{ij}
\eqno(69)
$$
     By extending the definition of $f_{ij}$ as
     $f_{ij}=f_{i\overline j}=f_{\overline i j}=f_{\overline i\overline j}$ we obtain
$$
\Tr_c( \ad_i\ad_j q_{il}q_{jk}a_ka_l \HR\HW)= \tr_c(\HR\HW)\sum_{i<j}\mu_i^{(r)}\mu_j^{(r)} f^{(2)}_{ij}
 [ {1\over 2}(\de_1)_{ii}(\de_1)_{jj}- (\de_1)_{ij}(\de_1)_{ji} ]
\eqno(70)
$$ 
     where now the sum is unrestricted, that is, it extends to all eigenvectors. The first term has the 
     form
     of  a direct contribution and the second of  an exchange contribution.
     As a final point, the matrix elements are of the type
$$
(\de_1)_{ij}= [ U^{-1} \left(\begin{array}{cc} q & 0\\ 0 & \tilde q \end{array}\right ) U ]_{ij}
\eqno(71)
$$
     where 
$$
U = \left(\begin{array}{cc} 1 & 0\\ 0 & R^*\end{array}\right ) V
\eqno(72)
$$
     where $V$ is the matrix that diagonalizes $M^{(r)}$, i,e,
$$
M^{(r)}= V \left(\begin{array}{cc} \mu^{(r)} & 0\\ 0 & \mu^{(r)}\end{array}\right ) V^{-1}
\eqno(73)
$$
     the $\mu^{(r)}$ being the eigenvalues.
\bigskip
\section {Conclusions.}
\bigskip
\par
     In this work we have discussed a variant of the thermal Hartree-Fock-Bogoliubov
     approximation (with the restoration of quantum numbers) whereby an approximation to the exact 
     free-energy is considered for a variational calculation which in the case of absence of the
     projectors gives the standard THFB. This variant is given directly in the canonical ensemble
     and equations for evaluation of traces, gradients of traces and one and two-body expectation values
     are given. 

\vfill
\eject
\end{document}